\documentclass[final,12pt,3p,times]{elsarticle}
\usepackage{amssymb,bbold}
\usepackage{amsmath}
\usepackage[utf8x]{inputenc}
\usepackage{natbib}
\usepackage{graphicx}
\usepackage{hyperref}
\usepackage{epstopdf}
\usepackage{pifont}
\usepackage{float}
\usepackage{ulem}
\usepackage{soul}
\usepackage{tikz}
\usepackage{caption}
\usepackage{orcidlink}
\graphicspath{{pdf/}}
\hypersetup{
  colorlinks=true,
  linkcolor=blue,
  citecolor=blue,
  filecolor=blue,
  urlcolor=blue,
  breaklinks=true
}

\biboptions{sort&compress}

\journal{Annals of Physics}
\RequirePackage{color}

\begin{document}

\begin{frontmatter}

\title{Quantum scattering in helically twisted geometries: Coulomb-like interaction and Aharonov-Bohm effect}

\author[ppgf]{Augusto Tadeu P. de Ara\'ujo}
\ead{augustofisico@gmail.com}

\author[ppgf,df]{Edilberto O. Silva\,\orcidlink{0000-0002-0297-5747}}
\ead{edilberto.silva@ufma.br}

\address[ppgf]{Programa de P\'{o}s-Gradua\c{c}\~{a}o em F\'{i}sica, Universidade Federal do Maranh\~{a}o, 65080-805, S\~{a}o Lu\'{i}s, Maranh\~{a}o, Brazil}
\address[df]{Coordena\c c\~ao do Curso de F\'{\i}sica -- Bacharelado, Universidade Federal do Maranh\~{a}o, 65085-580 S\~{a}o Lu\'{\i}s, Maranh\~{a}o, Brazil}

\begin{abstract}
We investigate the scattering of a charged quantum particle in a helically twisted background that induces an effective Coulomb-like interaction, in the presence of an Aharonov-Bohm (AB) flux. Starting from the nonrelativistic Schr\"odinger equation in the twisted metric, we derive the radial equation and show that, after including the AB potential, it can be mapped onto the same Kummer-type differential equation that governs the planar $2D$ Coulomb $+$ AB problem, with a geometry-induced Coulomb strength and the azimuthal quantum number shifted as $m\to m-\lambda$. We construct the exact scattering solutions, obtain closed expressions for the partial-wave $S$ matrix and phase shifts, and derive the corresponding scattering amplitude, differential cross section, and total cross section. We also show that the pole structure of the $S$ matrix is consistent with the bound-state quantization previously obtained for the helically twisted Coulomb-like problem.
\end{abstract}


\end{frontmatter}

\section{Introduction}

Geometry can act as an effective interaction in quantum mechanics.
Whenever the motion is restricted by constraints or by a nontrivial embedding, the kinetic term in curvilinear coordinates or in a curved background may generate additional scalar and vector structures that have no analogue in flat free space.
This idea has a long lineage, ranging from early formulations of quantum dynamics in curved spaces \citep{DeWitt1957} to modern constrained-quantization approaches \citep{JensenKoppe1971,daCosta1981}.
In the thin-layer quantization program, for instance, curvature produces a universal geometric potential on curved surfaces, with measurable consequences in mesoscopic rings, curved waveguides, and nanostructures \citep{daCosta1981,Encinosa2003,Atanasov2008,Atanasov2015}.
In parallel, the theory of geometric phases clarified that purely kinematical information, holonomy in parameter space, can imprint observable phases on wave functions \citep{Berry1984,ShapereWilczek1989,TomitaChiao1986}.
These two strands (geometry-induced forces and topology-induced phases) become particularly powerful in low-dimensional settings, where interference is robust and scattering can be dominated by long-range phase information rather than local forces.

A paradigmatic topological phase is the Aharonov--Bohm (AB) effect: a charged particle acquires a measurable phase shift even when the magnetic field vanishes along the particle trajectory, provided the vector potential has nontrivial flux \citep{EhrenbergSiday1949,AharonovBohm1959}.
After the first interference experiments \citep{Chambers1960}, the effect was demonstrated in the fully shielded-flux configuration by Tonomura \textit{et al.} \citep{Tonomura1986}, and it has become a cornerstone of mesoscopic physics and coherent transport \citep{Webb1985,OlariuPopescu1985,PeshkinTonomura1989}.
The AB phase also naturally connects to the broader framework of holonomies and geometric phases \citep{Berry1984,ShapereWilczek1989} and underlies fundamental developments such as anyonic statistics in two dimensions \citep{LeinaasMyrheim1977,Wilczek1982}.
From a scattering viewpoint, the AB interaction is notable because it is essentially topological and long-ranged, leading to subtle issues in partial-wave decompositions and in the definition of asymptotic states \citep{Ruijsenaars1983,Hagen1990}.
When additional long-range interactions are present, AB physics can interfere nontrivially with Coulomb-like phases, producing rich cross sections and analytic structures in the $S$ matrix.

A second major source of ``effective geometry'' arises from defects.
In the continuum description of crystalline media, dislocations and disclinations can be encoded in effective metrics and connections, providing an analogue-gravity setting in which curvature and torsion capture elastic distortions \citep{KatanaevVolovich1992,Moraes2000}.
Within the Einstein--Cartan viewpoint and its extensions, torsion couples to spin and can appear as a background field, with extensive reviews in the relativistic context \citep{Hehl1976,Kibble1976,Shapiro2002}.
In condensed matter, however, torsion can be realized effectively through defect densities or engineered structures, enabling controllable geometric effects on electronic and wave dynamics \citep{Moraes2000,KatanaevVolovich1992}.
This bridge has proven fruitful in graphene and related Dirac materials, where strain and lattice defects generate effective gauge fields and curvature/torsion-like terms \citep{CortijoVozmediano2007,deJuanCortijoVozmediano2010,Vozmediano2010,Guinea2010,Levy2010,Amorim2016}.
Beyond carbon systems, curved and twisted nanomembranes, rolled-up heterostructures, and helical photonic lattices have become mature platforms for geometry-driven wave manipulation \citep{Schmidt2001,Rechtsman2013,Strelow2018}.
Similarly, in ultracold atoms, spatial light modulators and digital micromirror devices (DMDs) enable time-dependent, reconfigurable potentials with fine control of trap geometry, providing a route to emulate nontrivial Laplace--Beltrami operators and synthetic gauge structures \citep{Gauthier2016}.

Motivated by these developments, helically twisted geometries provide a particularly appealing setting.
A helical twist can be viewed as a continuum model of a screw-like deformation, mixing angular and longitudinal coordinates in a way that resembles torsion-induced kinematics.
Recent analyses have shown that, even in a nominally free Schr\"odinger problem, the helical metric can induce an effective Coulomb-like $1/r$ interaction through the coupling between the twist parameter and conserved longitudinal and azimuthal momenta \citep{Azevedo2025}.
This geometry-induced Coulomb strength is not an external potential introduced by hand; it is a genuine kinematical consequence of the background.
Such a mechanism is conceptually analogous to other geometric interaction terms: curvature-induced scalar potentials \citep{daCosta1981,Atanasov2008}, defect-induced gauge-like couplings \citep{KatanaevVolovich1992,Moraes2000}, and strain-induced pseudo-fields in Dirac materials \citep{Vozmediano2010,Guinea2010,Levy2010}.
At the same time, the helical setting is experimentally suggestive because helical/twisted structures occur naturally and can be engineered in nanomembranes and photonic devices \citep{Schmidt2001,Rechtsman2013,Strelow2018}.

In this work, we ask a focused question: what happens when the geometry-induced Coulomb-like interaction generated by a helically twisted background is combined with a genuine topological AB flux?
This combination is natural for at least three reasons.
First, AB fluxes (or synthetic fluxes) are routinely invoked as control knobs in mesoscopic rings and waveguides \citep{Webb1985,OlariuPopescu1985,PeshkinTonomura1989}.
Second, long-range Coulomb phases have well-known signatures in scattering amplitudes and partial-wave phase shifts, and they imprint a characteristic pole structure in the $S$ matrix that encodes bound-state quantization.
Third, the planar two-dimensional Coulomb$+$AB problem is one of the few long-range scattering problems that remains analytically tractable: both bound and scattering sectors can be formulated in terms of confluent hypergeometric (Kummer) functions, and closed expressions for partial-wave $S$ matrices can be obtained \citep{Ruijsenaars1983,Nguyen2011EPJD}.
Therefore, if the helically twisted Coulomb-like problem with AB flux can be mapped onto the same Kummer-class equation, one gains an analytic handle on a geometry-modified scattering process with clear topological control.

Indeed, by starting from the Schr\"odinger equation in the helically twisted metric and minimally coupling an AB potential, we show that the resulting radial equation is equivalent, after transparent identifications, to the radial equation of the planar $2D$ Coulomb$+$AB problem \citep{Nguyen2011EPJD}.
The mapping is physically illuminating: the AB flux shifts the azimuthal quantum number, $m \to m-\lambda$, as expected from gauge invariance \citep{AharonovBohm1959,OlariuPopescu1985}, while the helical twist induces an effective Coulomb parameter proportional to $\omega k (m-\lambda)$, i.e., a geometry-controlled coupling that depends on both longitudinal momentum and the AB-shifted angular momentum.
This dependence is qualitatively distinct from conventional Coulomb scattering and produces channel-dependent phase shifts even at fixed energy.
Moreover, the $S$-matrix poles obtained from the analytic continuation of the scattering solutions are consistent with the bound-state quantization previously derived for the helical geometry without flux \citep{Azevedo2025}, now generalized by the AB shift.
Thus, bound and scattering sectors are unified at the level of the analytic structure of $S_m^{(\lambda)}$.

From the perspective of defect/geometry analogies, our results also provide a clean example of how a purely geometric background can emulate a long-range interaction and how that interaction interferes with a topological holonomy.
This speaks to a broad program: using controlled geometries (twist, curvature, disclinations/dislocations) to ``program'' effective forces and phases in quantum systems \citep{Moraes2000,KatanaevVolovich1992,Vozmediano2010,Amorim2016,Rechtsman2013}.
In particular, the helical twist plays a role analogous to a torsion-like coupling in effective theories, while the AB flux supplies a tunable topological phase.
Such interplay is relevant not only to electronic scattering, but also to wave transport in engineered helical media, where synthetic gauge fields and boundary conditions can mimic AB-like shifts \citep{Rechtsman2013,Gauthier2016}.

This paper is organized as follows.
In Sec. \ref{se}, we present the helically twisted metric, derive the Laplace--Beltrami operator with AB minimal coupling, and obtain the separated radial equation.
In Sec. \ref{scattering}, we construct the exact scattering solutions, derive closed expressions for the partial-wave $S$ matrix and phase shifts, and analyze the pole structure that connects the continuum sector to the bound-state spectrum.
We then build the scattering amplitude, the differential cross section, and the total cross section through a partial-wave expansion, and discuss the resulting angular patterns as functions of the twist and flux parameters.
Finally, in Sec. \ref{conclusion}, we summarize the main physical consequences and outline extensions, including spinful and relativistic generalizations, in which torsion and AB phases are expected to produce additional structure \citep{Hehl1976,Shapiro2002}.

\section{Schr\"odinger equation in the helically twisted geometry with AB flux \label{se}}

\subsection{Helically twisted metric}

We consider a quantum particle embedded in a three-dimensional helically twisted space characterized by a constant torsion parameter $\omega$. The spatial line element is taken as \cite{AoP.2026.484.170295}
\begin{equation}
ds^{2}
= dr^{2} + r^{2}\,d\varphi^{2} + \bigl(dz + \omega r\,d\varphi\bigr)^{2},
\label{metric-helical}
\end{equation}
where $(r,\varphi,z)$ are cylindrical coordinates and $\omega$ is a dimensionless parameter that controls the helical twisting. Expanding Eq.~\eqref{metric-helical}, we obtain
\begin{equation}
ds^{2}
= dr^{2} + r^{2}(1+\omega^{2})\,d\varphi^{2}
+ 2\,\omega r\,d\varphi\,dz + dz^{2}.
\label{metric-expanded}
\end{equation}
The corresponding spatial metric $g_{ij}$ and its inverse $g^{ij}$, written in the coordinate basis $(r,\varphi,z)$, read
\begin{equation}
g_{ij}
=
\begin{pmatrix}
1 & 0 & 0 \\[2pt]
0 & r^{2}(1+\omega^{2}) & \omega r \\[2pt]
0 & \omega r & 1
\end{pmatrix},
\qquad
\det g_{ij} = g = r^{2},
\label{metric-gij}
\end{equation}
\begin{equation}
g^{ij}
=
\begin{pmatrix}
1 & 0 & 0 \\[2pt]
0 & \dfrac{1}{r^{2}} & -\,\dfrac{\omega}{r} \\[6pt]
0 & -\,\dfrac{\omega}{r} & 1+\omega^{2}
\end{pmatrix}.
\label{metric-gij-inverse}
\end{equation}
Thus the volume element associated with the spatial geometry is $\sqrt{g}\,d^{3}x = r\,dr\,d\varphi\,dz$. In the limit $\omega\to 0$, the metric reduces to the flat cylindrical form $ds^{2}=dr^{2}+r^{2}d\varphi^{2}+dz^{2}$, and the helical twisting is switched off.

\subsection{Laplace-Beltrami operator and AB coupling}

The time-independent Schr\"odinger equation in this curved space for a particle of effective mass $\mu$ and electric charge $q$ reads
\begin{equation}
E\,\Psi(\mathbf{x})
= -\,\frac{\hbar^{2}}{2\mu}\,
\frac{1}{\sqrt{g}}\,
D_{i}\!\left(\sqrt{g}\,g^{ij} D_{j}\right)\Psi(\mathbf{x}),
\label{schrodinger-curved}
\end{equation}
where the gauge-covariant derivative is defined as
\begin{equation}
D_{i} = \partial_{i} - i\,\frac{q}{\hbar}A_{i}.
\label{covariant-derivative-AB}
\end{equation}
In the absence of external gauge fields, Eq.~\eqref{schrodinger-curved} reduces to the Laplace-Beltrami problem studied in Ref.~\cite{AoP.2026.484.170295}, where the helical geometry alone induces an effective Coulomb-like interaction. Here we further include an Aharonov-Bohm (AB) flux $\Phi$ along the $z$ axis.

In cylindrical coordinates, we choose the AB potential in the usual form
\begin{equation}
A_{r} = 0,
\qquad
A_{\varphi} = \frac{\Phi}{2\pi r},
\qquad
A_{z} = 0,
\label{AB-potential}
\end{equation}
so that the dimensionless flux parameter is
\begin{equation}
\lambda = \frac{q\Phi}{2\pi\hbar}
        = \frac{\Phi}{\Phi_{0}},
\qquad
\Phi_{0} = \frac{2\pi\hbar}{q}.
\label{AB-lambda}
\end{equation}
With this choice, the only component of the covariant derivative that is modified by the AB flux is
\begin{equation}
D_{\varphi} = \partial_{\varphi} - i\,\lambda,
\label{Dphi-AB}
\end{equation}
while $D_{r} = \partial_{r}$ and $D_{z} = \partial_{z}$.

Using Eqs.~\eqref{metric-gij-inverse}, \eqref{covariant-derivative-AB} and \eqref{Dphi-AB}, together with $\sqrt{g}=r$, the differential operator in Eq.~\eqref{schrodinger-curved} becomes
\begin{align}
\frac{1}{\sqrt{g}} D_{i}\!\left(\sqrt{g}\,g^{ij} D_{j}\right)
&= \frac{1}{r}\,\partial_{r}\!\left(r\,\partial_{r}\right)
+ \frac{1}{r^{2}}\bigl(D_{\varphi}\bigr)^{2}
- \frac{2\omega}{r}\,D_{\varphi}\partial_{z}
+ (1+\omega^{2})\,\partial_{z}^{2}.
\label{laplace-beltrami-AB}
\end{align}
Therefore, the Schr\"odinger equation with AB flux in the helically twisted space reads
\begin{equation}
E\,\Psi(r,\varphi,z)
= -\,\frac{\hbar^{2}}{2\mu}
\left[
\frac{1}{r}\,\partial_{r}\!\left(r\,\partial_{r}\right)
+ \frac{1}{r^{2}}\bigl(\partial_{\varphi}-i\lambda\bigr)^{2}
- \frac{2\omega}{r}\bigl(\partial_{\varphi}-i\lambda\bigr)\partial_{z}
+ (1+\omega^{2})\,\partial_{z}^{2}
\right]\Psi(r,\varphi,z).
\label{schrodinger-full-AB}
\end{equation}
In the limit $\lambda\to 0$, we recover the torsion-only case studied previously in Ref.~\cite{AoP.2026.484.170295}.

\subsection{Separation of variables and radial equation}

Because the metric~\eqref{metric-helical} is independent of $\varphi$ and $z$, and the AB potential~\eqref{AB-potential} preserves azimuthal and translational invariance, we can separate variables as
\begin{equation}
\Psi(r,\varphi,z)
= \frac{1}{\sqrt{2\pi L}}\,
e^{i m \varphi}\,
e^{i k z}\,
\psi_{m}(r),
\label{ansatz-separated}
\end{equation}
where $m\in\mathbb{Z}$ is the azimuthal quantum number, $k$ is the longitudinal wave number, and $L$ is a normalization length along the $z$ direction. Substituting the ansatz~\eqref{ansatz-separated} into Eq.~\eqref{schrodinger-full-AB} and using $D_{\varphi} \Psi = i(m-\lambda)\,\Psi$ and $\partial_{z}\Psi = i k\,\Psi$, we find the radial equation
\begin{equation}
E\,\psi_{m}(r)
= -\,\frac{\hbar^{2}}{2\mu}
\left[
\frac{1}{r}\,\frac{d}{dr}\!\left(r\,\frac{d}{dr}\right)
- \frac{(m-\lambda)^{2}}{r^{2}}
+ \frac{2\omega k (m-\lambda)}{r}
- (1+\omega^{2})k^{2}
\right]\psi_{m}(r).
\label{radial-eq-psi}
\end{equation}
It is convenient to remove the first derivative by defining $\psi_{m}(r) = r^{-1/2}\,f_{m}(r)$, which leads to a one-dimensional Schr\"odinger-like equation
\begin{equation}
-\,\frac{\hbar^{2}}{2\mu}\,\frac{d^{2}f_{m}}{dr^{2}}
+ V_{\text{eff}}(r)\,f_{m}(r)
= E\,f_{m}(r),
\label{schrodinger-1d}
\end{equation}
with the effective potential
\begin{equation}
V_{\text{eff}}(r)
= -\,\frac{\hbar^{2}}{2\mu}
\left[
-\,\frac{(m-\lambda)^{2}-\tfrac{1}{4}}{r^{2}}
+ \frac{2\omega k (m-\lambda)}{r}
- (1+\omega^{2})k^{2}
\right].
\label{Veff-AB}
\end{equation}
The first term in Eq.~\eqref{Veff-AB} represents the modified centrifugal barrier, shifted by $-1/(4r^{2})$ due to the transformation; the second term is an attractive (or repulsive) Coulomb-like contribution $\propto 1/r$ generated by the interplay between torsion and the angular and longitudinal momenta, and now explicitly dependent on the AB parameter $\lambda$; the last term is a constant energy shift associated with the torsion and longitudinal motion.

For $\lambda=0$, Eqs.~\eqref{radial-eq-psi} and \eqref{Veff-AB} reduce precisely to the radial equation and effective potential analyzed in Ref.~\cite{AoP.2026.484.170295}, confirming that the AB flux only shifts the azimuthal quantum number $m\to m-\lambda$, without altering the geometric origin of the Coulomb-like term. In what follows, we use Eq.~\eqref{radial-eq-psi} as the starting point to construct the scattering states and their phase shifts.

\section{Scattering states in the helically twisted Coulomb-Aharonov-Bohm system \label{scattering}}

In this section, we focus on the scattering regime of the geometry-induced Coulomb problem in the helically twisted background with AB flux. Our strategy is to reformulate the radial equation~\eqref{radial-eq-psi} in a way that makes its correspondence with the ``2D Coulomb + Aharonov-Bohm'' problem of Ref.~\cite{Nguyen2011EPJD} explicit, and then follow the same analytical steps used there to obtain the scattering wave functions, phase shifts, and $S$-matrix.

\subsection{Radial equation in the continuum sector}

Starting from Eq.~\eqref{radial-eq-psi}, we first bring all terms to the left-hand side and divide by $-\hbar^{2}/(2\mu)$. Following Ref.~\cite{AoP.2026.484.170295}, it is convenient to introduce the shifted continuum threshold
\begin{equation}
 E_{\infty} = -\frac{\hbar^{2}k^{2}}{2\mu}(1+\omega^{2}),
 \label{E-infty}
\end{equation}
which coincides with the asymptotic limit of the bound-state spectrum in the helically twisted Coulomb-like problem. We then parametrize the total energy as
\begin{equation}
 E = E_{\infty} + \frac{\hbar^{2}\kappa^{2}}{2\mu},
 \qquad \kappa>0,
 \label{E-scatt-param}
\end{equation}
so that $\kappa$ plays the role of the radial scattering wave number measured from the shifted continuum edge.

Substituting Eqs.~\eqref{E-infty} and \eqref{E-scatt-param} into the rearranged form of Eq.~\eqref{radial-eq-psi}, the constant terms combine to give $\kappa^{2}$, and the radial equation simplifies to
\begin{equation}
 \frac{d^{2}\psi_{m}}{dr^{2}} + \frac{1}{r}\frac{d\psi_{m}}{dr}
 + \left[
 \kappa^{2}
 - \frac{(m-\lambda)^{2}}{r^{2}}
 + \frac{2\omega k (m-\lambda)}{r}
 \right]\psi_{m}(r) = 0.
 \label{radial-AB-kappa-2}
\end{equation}
This equation clearly exhibits the structure of a two-dimensional Coulomb-like problem with an effective angular momentum $|m-\lambda|$ and a Coulomb parameter proportional to $\omega k (m-\lambda)$.

\subsection{Mapping to the 2D Coulomb + Aharonov-Bohm form}

The radial equation for the ``2D Coulomb + Aharonov-Bohm'' problem analysed in Ref.~\cite{Nguyen2011EPJD} can be written as
\begin{equation}
 \frac{d^{2}R^{(\lambda)}}{dr^{2}} + \frac{1}{r}\frac{dR^{(\lambda)}}{dr}
 + \left[
 k^{2}
 - \frac{|m+\lambda|^{2}}{r^{2}}
 - \frac{2\beta}{r}
 \right] R^{(\lambda)}(r) = 0,
 \label{Nguyen-radial-kbeta}
\end{equation}
with $k^{2} = 2ME/\hbar^{2}$ and $\beta = M\alpha/\hbar^{2}$, where $\alpha$ is the Coulomb coupling. Comparing Eq.~\eqref{Nguyen-radial-kbeta} with our Eq.~\eqref{radial-AB-kappa-2}, we see that the two problems are mathematically equivalent in the scattering sector, provided we make the identifications
\begin{equation}
 \kappa^{2} \;\longleftrightarrow\; k^{2},
 \qquad
 |m-\lambda| \;\longleftrightarrow\; |m+\lambda|,
 \qquad
 -2\beta_{\mathrm{geom}} \;\longleftrightarrow\; 2\omega k (m-\lambda),
 \label{identifications}
\end{equation}
where we have introduced the geometry-induced Coulomb-like coupling
\begin{equation}
 \beta_{\mathrm{geom}} \equiv -\omega k (m-\lambda).
 \label{beta-geom}
\end{equation}
Equivalently, in terms of a Coulomb-like strength $\alpha_{\mathrm{geom}}$, we can write
\begin{equation}
 \alpha_{\mathrm{geom}} = -\frac{\hbar^{2}}{\mu}\,\omega k (m-\lambda).
 \label{alpha-geom}
\end{equation}
Thus, the helically twisted space with torsion parameter $\omega$ and AB flux $\lambda$ realizes an effective ``2D Coulomb + Aharonov-Bohm'' problem in which the Coulomb charge is replaced by a geometry-induced coupling proportional to $\omega k (m-\lambda)$, while the azimuthal quantum number is shifted only by the AB flux and is not modified by the geometry itself.

With the identifications~\eqref{identifications}--\eqref{alpha-geom}, the scattering sector of our problem is governed by the same Kummer-type differential equation as in Ref.~\cite{Nguyen2011EPJD}. Following the standard procedure, we introduce the dimensionless variable $\xi = 2i\kappa r$ and write the radial wave function as
\begin{equation}
 \psi_{m}(r) = e^{i\kappa r} (2\kappa r)^{|m-\lambda|}\,W_{m}(\xi),
 \label{psi-ansatz-scatt}
\end{equation}
which incorporates the correct behavior near the origin and the oscillatory character at infinity. Substitution of~\eqref{psi-ansatz-scatt} into Eq.~\eqref{radial-AB-kappa-2} leads to the confluent hypergeometric equation
\begin{equation}
 \xi W_{m}'' + \bigl[2|m-\lambda|+1 - \xi\bigr] W_{m}'
 - \left( \frac{|m-\lambda|+1}{2} - i\frac{\beta_{\mathrm{geom}}}{2\kappa} \right) W_{m} = 0,
 \label{Kummer-geom}
\end{equation}
whose regular solution at $\xi=0$ is given by
\begin{equation}
 W_{m}(\xi) = C^{(\lambda)}_{\kappa m}\,
 {}_{1}F_{1}\!\left(
 \frac{|m-\lambda|+1}{2} - i\frac{\beta_{\mathrm{geom}}}{2\kappa},
 2|m-\lambda|+1;\,\xi
 \right),
 \label{W-solution}
\end{equation}
where $C^{(\lambda)}_{\kappa m}$ is a normalization factor. Therefore, the radial scattering wave functions in the helically twisted Coulomb-Aharonov-Bohm problem read
\begin{equation}
 \psi_{\kappa m}^{(\lambda)}(r)
 = C^{(\lambda)}_{\kappa m}\,
 e^{i\kappa r} (2\kappa r)^{|m-\lambda|}
 {}_{1}F_{1}\!\left(
 \frac{|m-\lambda|+1}{2} - i\frac{\beta_{\mathrm{geom}}}{2\kappa},
 2|m-\lambda|+1;\,2i\kappa r
 \right),
 \label{psi-final-scatt}
\end{equation}
which is the direct geometric analogue of the scattering solutions obtained by Nguyen \textit{et al.} in Ref.~\cite{Nguyen2011EPJD}, with the torsion-induced coupling $\beta_{\mathrm{geom}}$ playing the role of the Coulomb parameter.

In the next subsections, we shall use Eq.~\eqref{psi-final-scatt} to extract the partial-wave phase shifts, construct the $S$-matrix, and analyze the scattering amplitude and differential cross section in the helically twisted geometry.

\subsection{Asymptotic behaviour and partial-wave phase shifts}
\label{subsec:asymptotic-phases}

In order to extract the partial-wave phase shifts, we must analyze the large-$r$ behaviour of the radial wave function given in Eq.~\eqref{psi-final-scatt}. For later convenience, we introduce the shorthand
\begin{equation}
 a_m \equiv \frac{|m-\lambda|+1}{2}
            - i\,\frac{\beta_{\mathrm{geom}}}{2\kappa},
 \qquad
 b_m \equiv 2|m-\lambda|+1.
 \label{ab-def}
\end{equation}
The overall constant $C^{(\lambda)}_{\kappa m}$ is an $m$-dependent normalization factor that drops out of the ratio defining the $S$-matrix.

\subsubsection*{Step 1: Asymptotic form of the confluent hypergeometric function}

For large argument $|\xi|\to\infty$, with $|\arg\xi|<\pi$, the confluent hypergeometric function has the standard asymptotic expansion
\begin{equation}
 {}_{1}F_{1}(a;b;\xi)
 \sim
 \frac{\Gamma(b)}{\Gamma(a)}\,e^{\xi}\,\xi^{a-b}
 \;+\;
 \frac{\Gamma(b)}{\Gamma(b-a)}\,e^{i\pi a}\,\xi^{-a}.
 \label{1F1-asymptotic-ext}
\end{equation}
In our case, $\xi = 2i\kappa r$, $a = a_m$, and $b = b_m$. Substituting these into Eq.~\eqref{1F1-asymptotic-ext} and combining the powers of $(2\kappa r)$ with the prefactor $(2\kappa r)^{|m-\lambda|}$ from Eq.~\eqref{psi-final-scatt}, one can verify that both contributions behave in modulus as $r^{-1/2}$ when $r\to\infty$, as expected for two-dimensional scattering.

\subsubsection*{Step 2: Large-$r$ form of the radial wave function}

After separating the phase factors and collecting all pieces, the asymptotic form of $\psi_{\kappa m}^{(\lambda)}(r)$ can be written as
\begin{align}
 \psi_{\kappa m}^{(\lambda)}(r)
 &\sim
 C^{(\lambda)}_{\kappa m}
 \biggl\{
 \mathcal{A}_{m}(\kappa)\,
 e^{3i\kappa r}\,
 (2\kappa r)^{-\frac{1}{2}-i\frac{\beta_{\mathrm{geom}}}{2\kappa}}
 +
 \mathcal{B}_{m}(\kappa)\,
 e^{i\kappa r}
 (2\kappa r)^{-\frac{1}{2}+i\frac{\beta_{\mathrm{geom}}}{2\kappa}}
 \biggr\},
 \label{psi-asym-A-B}
\end{align}
where we have defined
\begin{align}
 \mathcal{A}_{m}(\kappa)
 &\equiv
 \frac{\Gamma(b_m)}{\Gamma(a_m)}\,
 e^{i\frac{\pi}{2}(a_m-b_m)},
 \label{Am-def}
 \\[4pt]
 \mathcal{B}_{m}(\kappa)
 &\equiv
 \frac{\Gamma(b_m)}{\Gamma(b_m-a_m)}\,
 e^{i\frac{\pi}{2}a_m}.
 \label{Bm-def}
\end{align}
The two terms in Eq.~\eqref{psi-asym-A-B} represent two independent asymptotic solutions: one corresponding to an outgoing wave and the other to an incoming wave, with amplitudes proportional to $\mathcal{A}_{m}(\kappa)$ and $\mathcal{B}_{m}(\kappa)$, respectively.

As usual in two dimensions, we can write the asymptotic behaviour in the symbolic form
\begin{equation}
 \psi_{\kappa m}^{(\lambda)}(r)
 \xrightarrow[r\to\infty]{}
 A_{m}^{(\lambda)}(\kappa)\,
 \frac{e^{-i\kappa r}}{\sqrt{r}}
 +
 B_{m}^{(\lambda)}(\kappa)\,
 \frac{e^{+i\kappa r}}{\sqrt{r}},
 \label{psi-asym-pwaves}
\end{equation}
where $A_{m}^{(\lambda)}$ and $B_{m}^{(\lambda)}$ are linear combinations of $\mathcal{A}_{m}$ and $\mathcal{B}_{m}$ (and of the common factor $C^{(\lambda)}_{\kappa m}$). The detailed expressions of these combinations are not needed because the physically relevant quantity is the ratio of outgoing to incoming amplitudes.

\subsubsection*{Step 3: Partial-wave S-matrix and phase shifts}

By definition, the partial-wave $S$-matrix element is given by the ratio between the outgoing and incoming amplitudes,
\begin{equation}
 S_{m}^{(\lambda)}(\kappa)
 = e^{2i\delta_{m}^{(\lambda)}(\kappa)}
 = \frac{B_{m}^{(\lambda)}(\kappa)}{A_{m}^{(\lambda)}(\kappa)}.
 \label{Sm-def-AB-ext}
\end{equation}
By explicitly tracking the phase factors in Eq.~\eqref{psi-asym-A-B} and comparing with the standard decomposition in Bessel and Hankel functions (see, for instance, the detailed analysis in Ref.~\cite{Nguyen2011EPJD}), one finds that the result can be cast in the compact form
\begin{equation}
 S_{m}^{(\lambda)}
 = S_{m}^{\mathrm{(AB)}}\,
 \frac{
 \Gamma\!\left(|m-\lambda|+\tfrac{1}{2} + i\,\eta_{m}\right)
 }{
 \Gamma\!\left(|m-\lambda|+\tfrac{1}{2} - i\,\eta_{m}\right)
 },
 \qquad
 \eta_{m} \equiv \frac{\beta_{\mathrm{geom}}}{\kappa}
 = -\,\frac{\omega k (m-\lambda)}{\kappa},
 \label{Sm-final-extended}
\end{equation}
where
\begin{equation}
 S_{m}^{\mathrm{(AB)}} = \exp\!\left[i\pi\bigl(|m| - |m-\lambda|\bigr)\right]
 \label{Sm-AB-extended}
\end{equation}
is exactly the $S$-matrix of the pure Aharonov-Bohm problem (obtained by taking $\omega=0$, i.e., $\beta_{\mathrm{geom}}=0$).

From Eq.~\eqref{Sm-final-extended}, the total phase shift can be written as
\begin{equation}
 \delta_{m}^{(\lambda)}
 = \delta_{m}^{\mathrm{(AB)}}
 + \delta_{m}^{\mathrm{(geom)}},
 \label{delta-split-extended}
\end{equation}
with
\begin{equation}
 \delta_{m}^{\mathrm{(AB)}} = \frac{\pi}{2}\bigl(|m| - |m-\lambda|\bigr),
 \label{delta-AB-extended}
\end{equation}
which is the purely topological contribution of the AB flux, and
\begin{equation}
 \delta_{m}^{\mathrm{(geom)}}
 = \arg\Gamma\!\left(|m-\lambda|+\tfrac{1}{2} + i\,\eta_{m}\right),
 \label{delta-geom-extended}
\end{equation}
which is the geometric contribution associated with the torsion-induced Coulomb-like potential. This result shows explicitly that the effect of the helical geometry enters the $S$-matrix only through the dimensionless parameter $\eta_{m} = \beta_{\mathrm{geom}}/\kappa$, whereas the purely AB effect is encoded in the factor $S_{m}^{\mathrm{(AB)}}$ and in the shift $m\to m-\lambda$.

\begin{figure}[tbhp]
\centering
\includegraphics[scale=0.3]{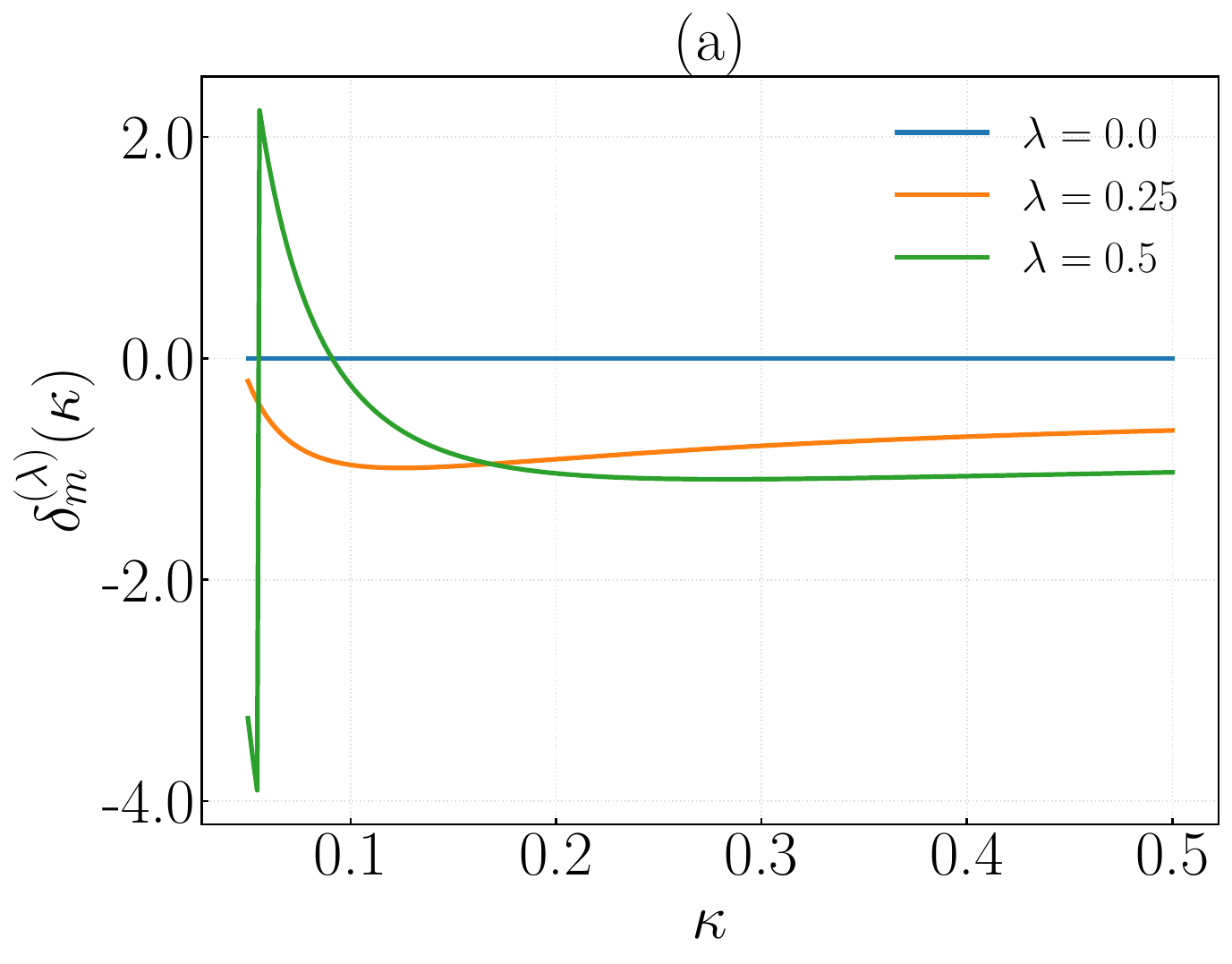}
\includegraphics[scale=0.3]{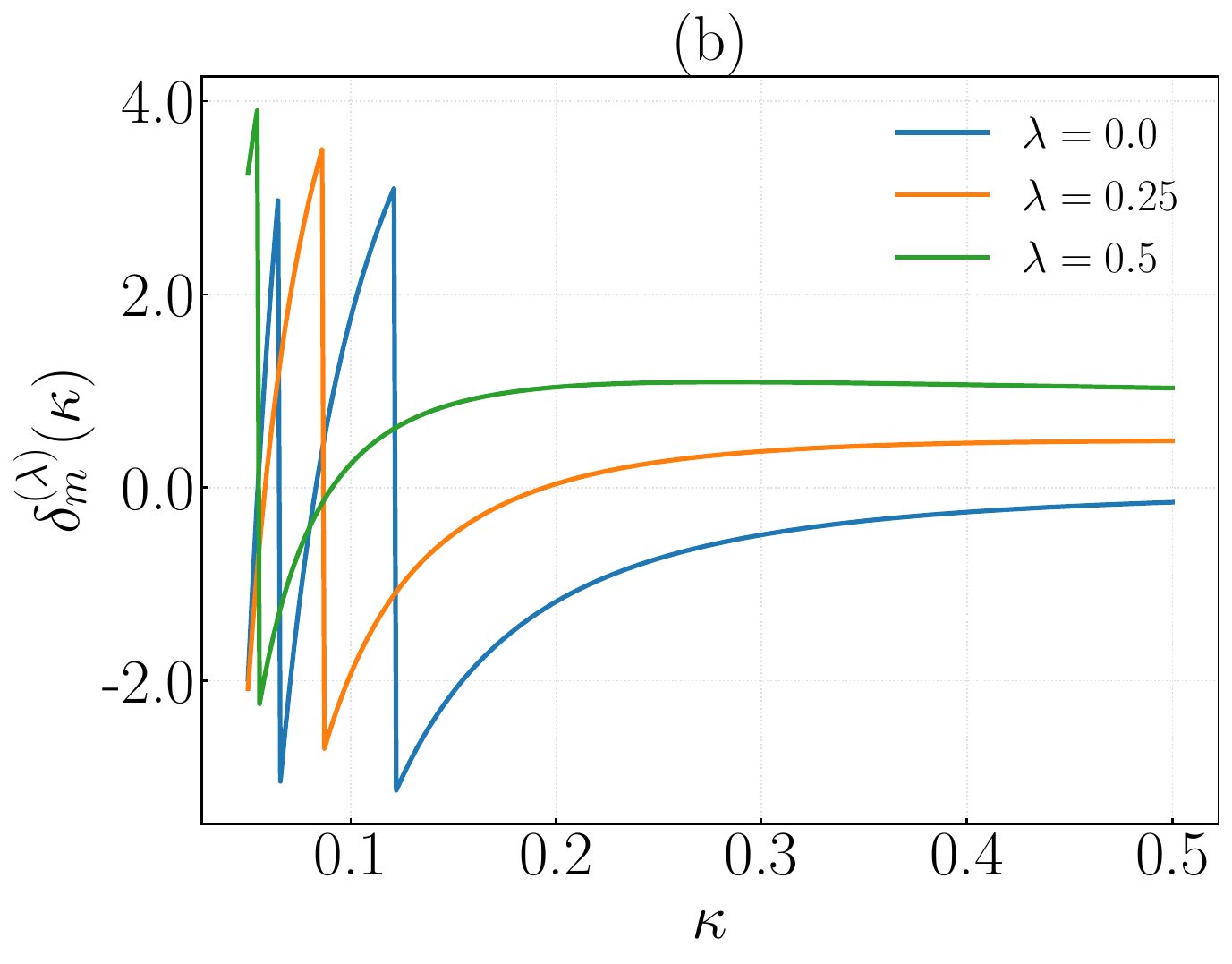}
\caption{Phase shift $\delta^{(\lambda)}_{m}(\kappa)$ as a function of the radial wave number $\kappa$ for $\omega=0.5$ and $k=1.0$. Each curve corresponds to a different value of the Aharonov--Bohm flux parameter, $\lambda=0$ (blue), $\lambda=0.25$ (orange), and $\lambda=0.5$ (green). Panel~(a) shows the $s$-wave channel $m=0$, while panel~(b) refers to the first angular channel $m=1$.}
\label{fig:delta_m0_m1}
\end{figure}
Figures~\ref{fig:delta_m0_m1}(a) and \ref{fig:delta_m0_m1}(b) display the total phase shift $\delta^{(\lambda)}_{m}(\kappa)$ as a function of the radial wave number for the lowest angular channels $m=0$ and $m=1$. For $m=0$, the phase shift starts from a large negative value at very small $\kappa$ and quickly saturates to a $\kappa$-independent plateau, whose height is shifted downward as the AB flux increases, reflecting the purely topological contribution $\delta^{(\mathrm{AB})}_{0}=-\tfrac{\pi}{2}|\lambda|$. For $m=1$, shown in panel~\ref{fig:delta_m0_m1}(b), the behaviour near threshold is more pronounced and the plateaus at larger $\kappa$ are shifted upwards by the AB term $\delta^{(\mathrm{AB})}_{1}$, while the geometry-induced Coulomb contribution $\delta^{(\mathrm{geom})}_{1}$ controls the smooth approach to these asymptotic values. Together, these plots illustrate how helical torsion and AB flux combine to deform the partial-wave phase shifts in a channel-dependent manner.
\paragraph{On the ``sawtooth'' behavior of the phase shift near $\kappa\simeq 0$}
The sharp ``sawtooth'' profile observed in Fig.~\ref{fig:delta_m0_m1} for $\kappa$ close to zero is a branch (phase-wrapping) effect rather than a new physical singularity. From Eq.~(\ref{delta-split-extended}), we have
$\delta^{(\lambda)}_{m}(\kappa)=\delta^{(\mathrm{AB})}_{m}+\delta^{(\mathrm{geom})}_{m}$ with
$\delta^{(\mathrm{geom})}_{m}=\arg\Gamma\!\big(|m-\lambda|+\tfrac12+i\eta_m\big)$ and
$\eta_m=\beta_{\mathrm{geom}}/\kappa=-\omega k (m-\lambda)/\kappa$.
Hence $|\eta_m|\propto 1/\kappa$ diverges as $\kappa\to 0^+$ (in particular, for $m=0$ and $\lambda\neq 0$ one has $\eta_0=\omega k\lambda/\kappa$), implying a very rapid variation of $\arg\Gamma(\cdot)$ in the threshold region. For $|\eta|\gg 1$, the Stirling-type expansion yields
$\arg\Gamma(a+i\eta)=\Im\ln\Gamma(a+i\eta)\sim \eta\ln|\eta|-\eta+\mathrm{const}+O(1/\eta)$
(with $a=|m-\lambda|+\tfrac12$), so $\delta_m^{(\mathrm{geom})}$ changes quasi-linearly in $\eta$ and thus extremely fast in $\kappa$.
When the phase is computed as a principal value (e.g. in $(-\pi,\pi]$), the continuous growth of $\arg\Gamma$ is mapped into repeated $\pm 2\pi$ jumps, producing the apparent ``sawtooth'' pattern. A smooth representation is obtained by using an \textit{unwrapped} phase, namely by tracking $\Im\ln\Gamma(\cdot)$ continuously (e.g. via $\mathrm{log}\,\Gamma$) and adding/subtracting $2\pi$ whenever the principal value crosses $\pm\pi$.

\begin{figure}[tbhp]
\centering
\includegraphics[scale=0.5]{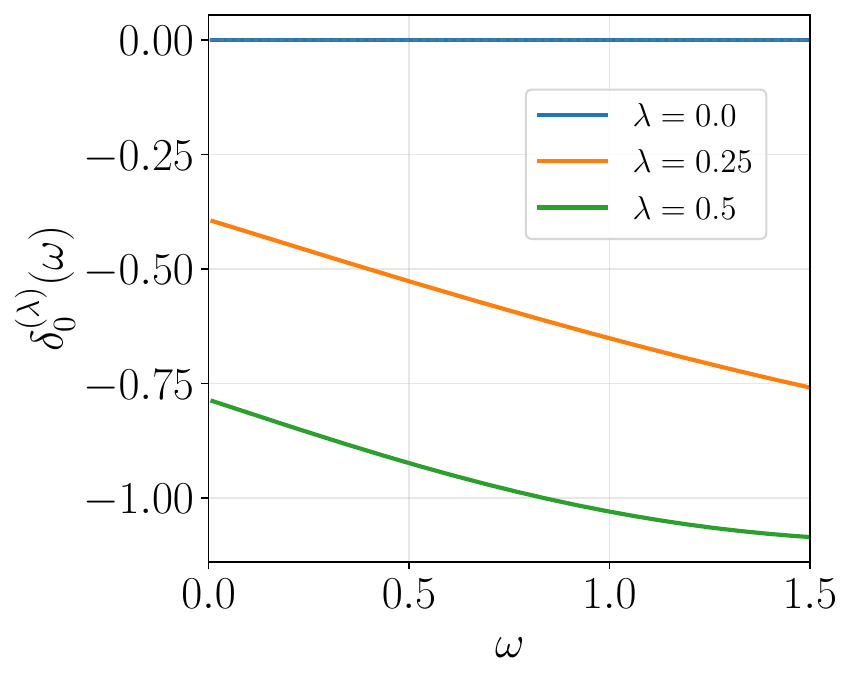}
\includegraphics[scale=0.5]{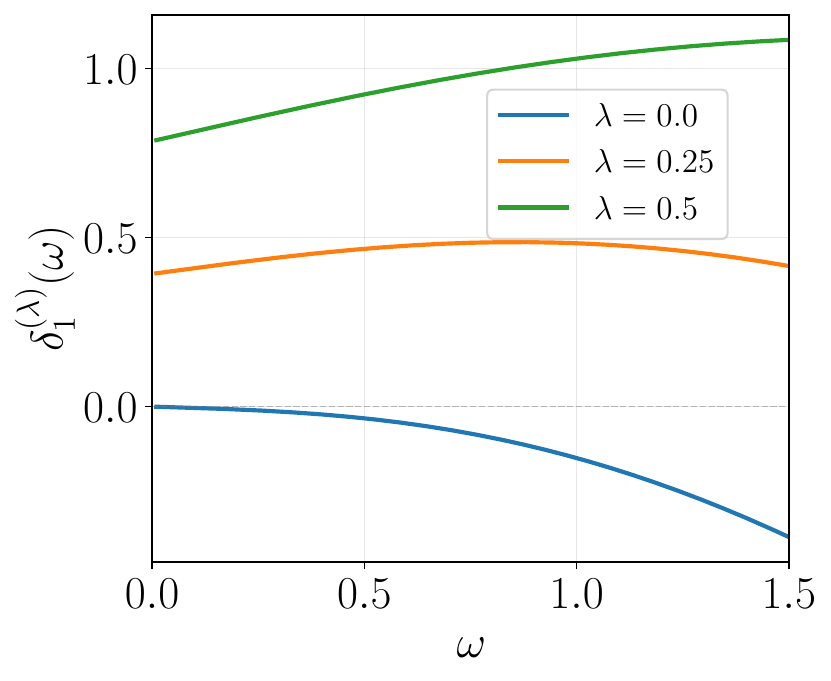}
\caption{Phase shift $\delta^{(\lambda)}_{m}(\omega)$ as a function of the torsion parameter $\omega$ for fixed $\kappa=1.0$ and $k=1.0$. Each curve corresponds to a different value of the Aharonov--Bohm flux parameter, $\lambda=0$ (blue), $\lambda=0.25$ (orange), and $\lambda=0.5$ (green). Panel~(a) shows the $s$-wave channel $m=0$, while panel~(b) refers to the first angular channel $m=1$.}
\label{fig:delta_vs_omega}
\end{figure}
To further elucidate the role of the helical geometry, Figs.~\ref{fig:delta_vs_omega}(a) and \ref{fig:delta_vs_omega}(b) display the total phase shift as a function of the torsion parameter $\omega$ for fixed $\kappa=1$ and $k=1$. For the $s$-wave channel $m=0$, shown in Fig. ~\ref{fig:delta_vs_omega}(a), the phase shift exhibits a monotonic dependence on $\omega$: in the absence of AB flux ($\lambda=0$), the phase shift vanishes identically since $\eta_0 = 0$ when $m=0$; however, for nonzero $\lambda$, the effective angular momentum $|m-\lambda|=|\lambda|$ becomes fractional and the geometry-induced Coulomb parameter $\eta_0 = \omega k \lambda/\kappa$ grows linearly with $\omega$, producing a steady increase in $|\delta_0^{(\lambda)}|$. The AB flux thus activates a geometric phase shift in the $s$-wave channel that would otherwise be absent. For the $m=1$ channel, shown in Fig.~\ref{fig:delta_vs_omega}(b), the phase shift grows more rapidly with $\omega$ due to the larger effective Coulomb coupling $\eta_1 \propto \omega(1-\lambda)$. The spread between curves at different $\lambda$ values increases with $\omega$, demonstrating that the interplay between torsion and AB flux becomes more pronounced in the strong-twist regime. These results confirm that the torsion parameter $\omega$ acts as a continuous control knob for the geometry-induced scattering phases, with the AB flux providing an additional handle to tune the effective interaction strength in each angular momentum channel.

\subsection{S-matrix poles and connection with bound states}
\label{subsec:Sm-poles}

Although we focus on the scattering sector, the analytic structure of the partial-wave $S$-matrix provides a direct link to the bound-state spectrum. In particular, the poles of $S_{m}^{(\lambda)}(E)$ in the complex energy plane encode the discrete energy levels of normalizable states, in close analogy to the standard Coulomb and Coulomb+AB problems.

The expression for $S_{m}^{(\lambda)}(\kappa)$ given in Eq.~\eqref{Sm-final-extended} shows that possible poles can only arise from the Gamma functions in the numerator or in the denominator. Physical bound states correspond to poles that appear when the scattering energy is analytically continued below the continuum threshold, such that the radial solution becomes normalizable. In the current parameterization, these poles correspond to the denominator of the Gamma function.

\subsubsection*{Step 1: Analytic continuation and the Coulomb parameter}

The partial-wave solutions in the scattering regime are written in terms of the real wave number $\kappa>0$. To connect with bound states, it is convenient to perform the usual analytic continuation
\begin{equation}
 \kappa \;\longrightarrow\; i\rho,
 \qquad
 \rho>0,
 \label{kappa-to-rho}
\end{equation}
where $\rho$ plays the role of the radial decay constant in the bound-state sector. Under this continuation, the Coulomb-like parameter $\eta_{m}(\kappa)$ becomes
\begin{equation}
 \eta_{m}(i\rho)
 = \frac{\beta_{\mathrm{geom}}}{i\rho}
 = -\,i\,\frac{\beta_{\mathrm{geom}}}{\rho},
 \label{eta-analytic}
\end{equation}
so that $-i\,\eta_{m}(i\rho) = \beta_{\mathrm{geom}}/\rho$ is real and plays exactly the same role as the ratio between the Coulomb-like coupling and the decay constant $\rho$ in the bound-state problem.

\subsubsection*{Step 2: Pole condition from the Gamma function}

The Gamma function $\Gamma(z)$ has simple poles at $z=0,-1,-2,\dots$. The poles of $S_{m}^{(\lambda)}$ associated with bound states come from the denominator Gamma function in Eq.~\eqref{Sm-final-extended}. The pole condition reads
\begin{equation}
 |m-\lambda|+\tfrac{1}{2} - i\,\eta_{m}(\kappa)
 = -n,
 \qquad
 n = 0,1,2,\dots,
 \label{pole-condition-general}
\end{equation}
which must be understood after analytic continuation to $\kappa=i\rho$ and to energies below the continuum threshold. Using Eq.~\eqref{eta-analytic}, this condition becomes
\begin{equation}
 |m-\lambda|+\tfrac{1}{2}
 + \frac{\beta_{\mathrm{geom}}}{\rho}
 = -n,
 \qquad
 n = 0,1,2,\dots.
 \label{pole-condition-rho}
\end{equation}

At this point, it is convenient to compare Eq.~\eqref{pole-condition-rho} with the bound-state quantization rule obtained in Ref.~\cite{AoP.2026.484.170295}. In that work, the regular solution at the origin is written in terms of a confluent hypergeometric function with parameter $a_{\text{bound}}(E,m,\omega,k) = \frac{1}{2} + |m| - \Xi(E,m,\omega,k)$, and normalizability requires $a_{\text{bound}}(E,m,\omega,k) = -n$ for $n=0,1,2,\dots$, which is Eq.~(16) of Ref.~\cite{AoP.2026.484.170295}. The quantity $\Xi(E,m,\omega,k)$ is proportional to $k m\omega/(\hbar\rho)$ and encodes the geometry-induced Coulomb coupling divided by the decay constant $\rho$.

Equation~\eqref{pole-condition-rho} has exactly the same structure, with two modifications: (i) the azimuthal quantum number is replaced by its AB-shifted version $m\to m-\lambda$; and (ii) the combination $\beta_{\mathrm{geom}}/\rho$ plays the role of $\Xi(E,m,\omega,k)$, i.e., it is proportional to $\omega k (m-\lambda)/\rho$. Therefore, Eq.~\eqref{pole-condition-rho} is equivalent to the bound-state quantization rule of Ref.~\cite{AoP.2026.484.170295}, now with the AB shift $m\to m-\lambda$.

\subsubsection*{Step 3: Consistency with the known bound-state spectrum}

Since Eq.~\eqref{pole-condition-rho} reproduces the same quantization condition as the one in Ref.~\cite{AoP.2026.484.170295}, the discrete energies that follow from the poles of $S_{m}^{(\lambda)}(E)$ coincide with the exact bound-state spectrum already obtained there, up to the replacement $m\to m-\lambda$ induced by the AB flux.\footnote{In particular, in the limit $\lambda\to 0$ the pole condition reduces to the quantization rule used in Ref.~\cite{AoP.2026.484.170295}, and the corresponding energies $E_{n,m}$ agree identically with the explicit expression reported there.} No new bound-state levels are generated by the scattering analysis; the S-matrix simply reorganizes the same information in terms of its analytic structure in the complex-energy plane.

It is important to emphasize that we do not need to rewrite the explicit expression for $E_{n,m}$ here. The full bound-state spectrum and wave functions in the helically twisted Coulomb-like problem have already been derived in closed form in Ref.~\cite{AoP.2026.484.170295}, and those results remain valid in the present setting after the substitution $m\to m-\lambda$ in the effective Coulomb coupling.

\subsubsection*{Step 4: Role of the S-matrix poles in the present work}

The discussion above shows that the S-matrix given in Eq.~\eqref{Sm-final-extended} provides a unified analytic description of both the scattering and the bound-state sectors of the helically twisted Coulomb-Aharonov-Bohm problem: the continuous spectrum is encoded in the phase of $S_{m}^{(\lambda)}$, while the discrete spectrum is encoded in its poles. In the present article, however, we are primarily interested in the scattering states and in the geometry- and flux-dependent phase shifts and cross sections. For this reason, we use the pole structure of $S_{m}^{(\lambda)}$ only as a consistency check of our scattering formalism and as a formal bridge connecting the scattering amplitudes to the bound-state spectrum previously obtained in Ref.~\cite{AoP.2026.484.170295}, without rederiving the latter in detail.

\subsection{Scattering amplitude and differential cross section}
\label{subsec:amplitude-cross-section}

Once the partial-wave $S$-matrix elements $S_{m}^{(\lambda)}$ are known, the scattering amplitude and the differential cross section follow from standard partial-wave methods. In this subsection, we make this connection explicit for the helically twisted Coulomb-Aharonov-Bohm system.

\subsubsection*{Step 1: Partial-wave expansion of the incident plane wave}

We consider an incident plane wave propagating along the $x$ axis in the transverse $(r,\theta)$ plane, with a fixed longitudinal momentum $k$ along the $z$ direction. Its spatial dependence in the plane is $e^{i\kappa x} = e^{i\kappa r\cos\theta}$, where $\kappa>0$ is the radial scattering wave number defined in Eq.~\eqref{E-scatt-param}, and $\theta$ is the polar angle measured from the $x$ axis. In terms of cylindrical harmonics, the plane wave can be expanded as
\begin{equation}
 e^{i\kappa r\cos\theta}
 = \sum_{m=-\infty}^{+\infty}
 i^{m}\,J_{m}(\kappa r)\,e^{im\theta},
 \label{plane-wave-expansion}
\end{equation}
where $J_{m}$ denotes the Bessel function of the first kind. For large $r$, the Bessel functions behave as
\begin{equation}
 J_{m}(\kappa r)
 \xrightarrow[r\to\infty]{}
 \sqrt{\frac{1}{2\pi\kappa r}}
 \left[
 e^{i\left(\kappa r - \frac{\pi m}{2} - \frac{\pi}{4}\right)}
 + e^{-i\left(\kappa r - \frac{\pi m}{2} - \frac{\pi}{4}\right)}
 \right],
 \label{Jm-asymptotic}
\end{equation}
which is a superposition of incoming and outgoing cylindrical waves with equal amplitude.

\subsubsection*{Step 2: Asymptotic form of the full wave function}

In the presence of the geometry-induced Coulomb-like interaction and the AB flux, the full stationary scattering state can be written as a partial-wave sum
\begin{equation}
 \Psi(r,\theta,z)
 = \frac{e^{ikz}}{\sqrt{2\pi L}}
 \sum_{m=-\infty}^{+\infty}
 a_{m}\,\psi_{\kappa m}^{(\lambda)}(r)\,e^{im\theta},
 \label{Psi-partialwave}
\end{equation}
where $\psi_{\kappa m}^{(\lambda)}(r)$ is the radial scattering solution given in Eq.~\eqref{psi-final-scatt} and $a_{m}$ are coefficients chosen such that, in the absence of scattering, the sum reproduces the incident plane wave. In practice, this normalization can be fixed by matching the incoming part to Eq.~\eqref{plane-wave-expansion} at large $r$.

Using the asymptotic behaviour given in Eq.~\eqref{psi-asym-pwaves} and ensuring that the incoming part matches the plane wave expansion, the full wave function can be cast in the standard scattering form
\begin{equation}
 \Psi(r,\theta,z)
 \xrightarrow[r\to\infty]{}
 e^{i\kappa r\cos\theta}\,e^{ikz}
 + f^{(\lambda)}(\theta)\,\frac{e^{i\kappa r}}{\sqrt{r}}\,
 e^{ikz},
 \label{Psi-asym-standard}
\end{equation}
where the first term is the incident plane wave and the second term defines the scattering amplitude $f^{(\lambda)}(\theta)$.

\subsubsection*{Step 3: Scattering amplitude in terms of $S_{m}^{(\lambda)}$}

Using the definition of the partial-wave $S$-matrix given in Eq.~\eqref{Sm-def-AB-ext} and the normalization that matches the incoming wave to Eq.~\eqref{plane-wave-expansion}, the scattering amplitude can be written as
\begin{equation}
 f^{(\lambda)}(\theta)
 = \frac{1}{\sqrt{2\pi\kappa}}
 \sum_{m=-\infty}^{+\infty}
 \left[S_{m}^{(\lambda)}(\kappa) - 1\right] e^{im\theta}.
 \label{f-theta-final}
\end{equation}
This is the standard result for two-dimensional scattering in the presence of an azimuthally symmetric potential, now with $S_{m}^{(\lambda)}$ given by the geometry- and flux-dependent expression in Eq.~\eqref{Sm-final-extended}.

\subsubsection*{Step 4: Differential cross section and limiting cases}

The differential cross section in the transverse plane is defined as
\begin{equation}
 \frac{d\sigma}{d\theta} = \bigl|f^{(\lambda)}(\theta)\bigr|^{2},
 \label{diff-cross-section-final}
\end{equation}
with $f^{(\lambda)}(\theta)$ given by Eq.~\eqref{f-theta-final}. The nontrivial dependence on the helical geometry and on the AB flux is encoded in the interference of the partial waves through the phase shifts given in Eqs.~\eqref{delta-split-extended}--\eqref{delta-geom-extended}.

Two important limiting cases are immediately recovered:

\begin{itemize}
 \item \textbf{Pure Aharonov-Bohm limit ($\omega=0$):}  
 In this case $\beta_{\mathrm{geom}}=0$ and $\eta_{m}=0$, so that $S_{m}^{(\lambda)} \to S_{m}^{\mathrm{(AB)}}$, and the scattering amplitude reduces to the well-known AB result, with a purely topological angular dependence and no bound states.

 \item \textbf{Pure geometric Coulomb-like limit ($\lambda=0$):}  
 When the AB flux is switched off ($\lambda=0$), the azimuthal quantum number remains $m$ and the $S$-matrix reduces to the ratio of Gamma functions with $\eta_{m} = -\,\omega k m/\kappa$. The scattering is then entirely controlled by the geometry-induced Coulomb-like interaction, and the corresponding differential cross section displays the characteristic forward divergence associated with long-range Coulomb potentials, now modulated by the torsion parameter $\omega$ and the longitudinal momentum $k$.
\end{itemize}

In the general case, both $\omega$ and $\lambda$ are nonzero, and the interplay between the topological AB phase and the geometry-induced Coulomb phase leads to a rich angular pattern in the differential cross section. In the next section, we illustrate these features with explicit plots of $d\sigma/d\theta$ for representative values of $(\omega,\lambda,m)$ and discuss their physical interpretation.

\begin{figure}[tbhp]
    \centering
    \includegraphics[width=0.48\textwidth]{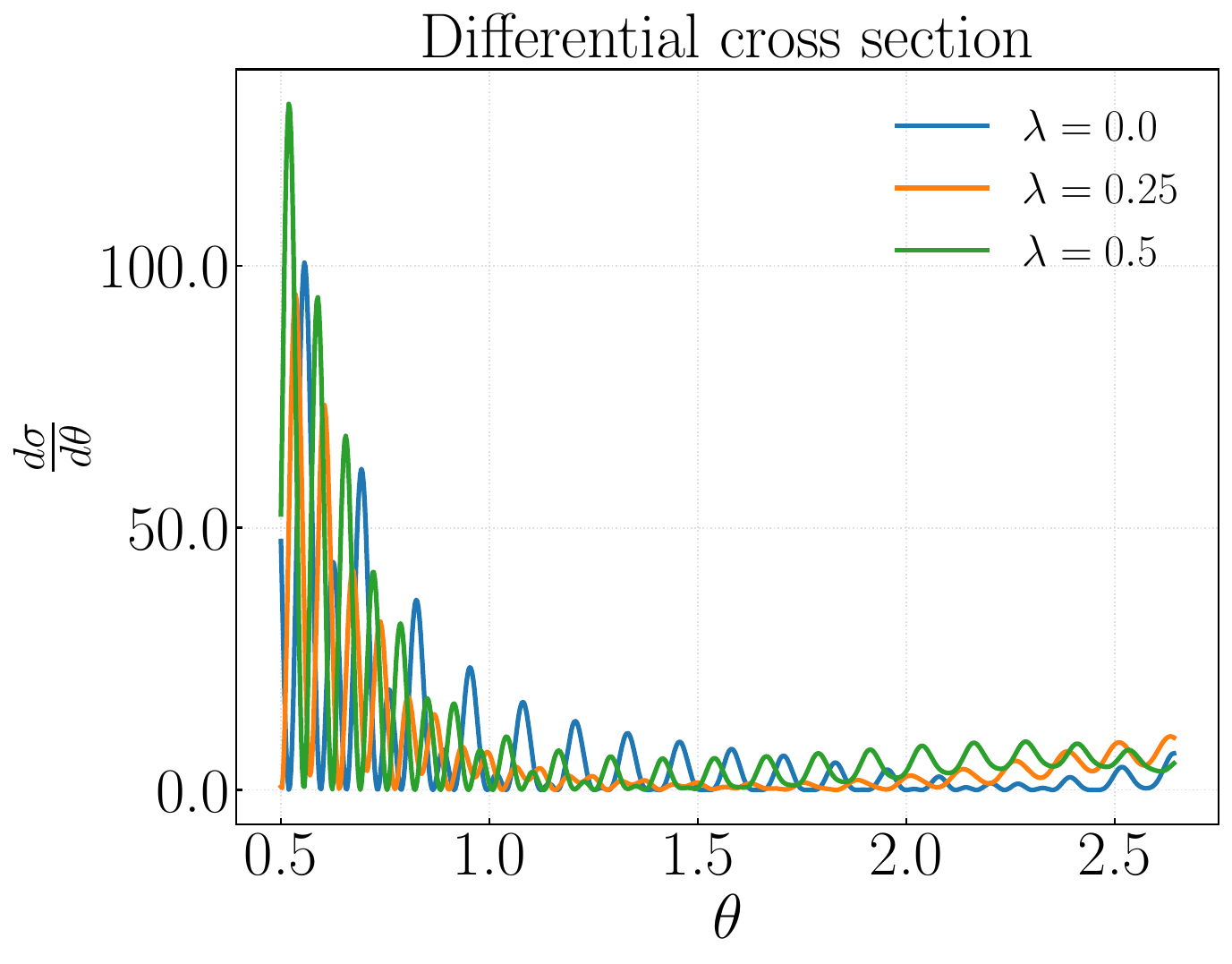}
    \caption{(Color online) Differential cross section 
    $d\sigma/d\theta$ as a function of the scattering angle 
    $\theta$ for the helically twisted Coulomb--Aharonov--Bohm 
    problem. The curves are evaluated for $\kappa = 1$, 
    $\omega = 0.7$, $k = 1$, and three values of the Aharonov--Bohm 
    parameter, $\lambda = 0.0$, $0.25$, and $0.5$, using the 
    partial--wave expansion with angular momenta in the range 
    $m = -50,\ldots,50$. The angular domain is restricted to 
    $0.5 \le \theta \le \pi - 0.5$ in order to exclude the 
    forward and backward singular directions.}
    \label{fig:dsigma_theta}
\end{figure}
Figure~\ref{fig:dsigma_theta} shows the differential cross section $d\sigma/d\theta$ as a function of the scattering angle $\theta$ for fixed $\kappa = 1$ and $\omega = 0.7$. For all values of the flux parameter, the cross section exhibits a pronounced maximum at small angles, followed by a damped oscillatory tail, which is the hallmark of interference among many partial waves in a long-range Coulomb-like interaction. Increasing the Aharonov--Bohm parameter from $\lambda = 0$ to $\lambda = 0.5$ enhances the small-angle peak and modifies both the amplitude and the phase of the oscillations at intermediate angles, revealing a strong flux control of the angular 
scattering pattern induced by the helical geometry.

\paragraph{Origin of the oscillatory angular pattern in $d\sigma/d\theta$}
The oscillatory (damped) profile displayed in Fig.~\ref{fig:dsigma_theta} follows directly from the partial-wave construction of the scattering amplitude in a long--range (Coulomb-like) problem combined with the AB-induced reshuffling of the angular channels. Indeed, the amplitude is obtained from the Fourier-like  (\ref{f-theta-final}) and Eq. (\ref{diff-cross-section-final}), so that the angular distribution is the result of coherent interference among many cylindrical harmonics.
For short--range potentials the phase shifts $\delta_m$ decay rapidly with $|m|$, and only a few partial waves contribute appreciably; by contrast, the geometry-induced Coulomb-like interaction produces slowly decaying (channel-dependent) phases, which makes a broad range of $m$ relevant and naturally generates an interference pattern with alternating maxima and minima as a function of $\theta$.
This same long-range character also explains the pronounced enhancement at small angles: forward peaking is a generic hallmark of Coulomb scattering and persists here in its geometry-induced analogue, motivating the exclusion of the immediate forward/backward directions in the plotted angular window.

The AB flux enters through the gauge-invariant shift $m\to m-\lambda$ in the radial problem and, consequently, in $S_{m}^{(\lambda)}$ and $\delta_{m}^{(\lambda)}$. This shift changes the relative phases among the contributing partial waves, thereby modifying both the amplitude and the \textit{phase} of the oscillations in $d\sigma/d\theta$ and allowing a flux-tunable control of the angular interference pattern.
Finally, from a numerical standpoint, Fig.~\ref{fig:dsigma_theta} is obtained by truncating the series in Eq. (\ref{f-theta-final}) to $|m|\le m_{\max}$.
Because long--range interactions require a comparatively large number of partial waves for convergence, the fine oscillatory texture at intermediate angles can show a mild dependence on $m_{\max}$ (a standard truncation/regularization effect in Fourier-like partial-wave sums), while the overall features, strong forward enhancement and a robust oscillatory tail, remain unchanged upon increasing $m_{\max}$ within the range used in our simulations.

\subsection{Total cross section}
\label{subsec:total-cross-section}

The total cross section provides a global measure of the scattering strength and is an important observable that complements the angular information contained in the differential cross section. In two-dimensional scattering, the total cross section can be obtained either by direct integration of the differential cross section or, more elegantly, via the optical theorem. In this subsection, we derive both approaches and discuss their implications for the helically twisted Coulomb-Aharonov-Bohm system.

\subsubsection*{Partial-wave expression for the total cross section}

The total cross section is defined as the integral of the differential cross section over all scattering angles,
\begin{equation}
 \sigma_{\mathrm{tot}} = \int_{0}^{2\pi} \frac{d\sigma}{d\theta}\,d\theta
 = \int_{0}^{2\pi} \bigl|f^{(\lambda)}(\theta)\bigr|^{2}\,d\theta.
 \label{sigma-tot-def}
\end{equation}
Substituting the partial-wave expansion of the scattering amplitude given in Eq.~\eqref{f-theta-final} and using the orthogonality of the angular harmonics,
\begin{equation}
 \int_{0}^{2\pi} e^{i(m-m')\theta}\,d\theta = 2\pi\,\delta_{mm'},
\end{equation}
we obtain the partial-wave decomposition of the total cross section,
\begin{equation}
 \sigma_{\mathrm{tot}} = \frac{1}{\kappa} \sum_{m=-\infty}^{+\infty} \bigl|S_{m}^{(\lambda)} - 1\bigr|^{2}.
 \label{sigma-tot-pw}
\end{equation}
Since the $S$-matrix elements are unimodular, $|S_{m}^{(\lambda)}|=1$, we can write
\begin{equation}
 \bigl|S_{m}^{(\lambda)} - 1\bigr|^{2} = \bigl|e^{2i\delta_{m}^{(\lambda)}} - 1\bigr|^{2}
 = 4\sin^{2}\delta_{m}^{(\lambda)},
 \label{Sm-minus-1-squared}
\end{equation}
and therefore
\begin{equation}
 \sigma_{\mathrm{tot}} = \frac{4}{\kappa} \sum_{m=-\infty}^{+\infty} \sin^{2}\delta_{m}^{(\lambda)}.
 \label{sigma-tot-sin2}
\end{equation}
This expression shows that the total cross section is determined by the sum of the squared sines of all partial-wave phase shifts. In the helically twisted Coulomb-AB problem, each phase shift $\delta_{m}^{(\lambda)}$ contains both the topological AB contribution and the geometry-induced Coulomb contribution, as given in Eqs.~\eqref{delta-split-extended}--\eqref{delta-geom-extended}.

\subsubsection*{Optical theorem}

An alternative, physically insightful route to the total cross section is provided by the optical theorem, which relates $\sigma_{\mathrm{tot}}$ to the imaginary part of the forward-scattering amplitude. In two dimensions, the optical theorem reads
\begin{equation}
 \sigma_{\mathrm{tot}} = \sqrt{\frac{8\pi}{\kappa}}\,\mathrm{Im}\!\left[e^{-i\pi/4}\,f^{(\lambda)}(0)\right].
 \label{optical-theorem}
\end{equation}
To verify this relation, we evaluate the forward scattering amplitude from Eq.~\eqref{f-theta-final},
\begin{equation}
 f^{(\lambda)}(0) = \frac{1}{\sqrt{2\pi\kappa}} \sum_{m=-\infty}^{+\infty} \left[S_{m}^{(\lambda)} - 1\right],
\end{equation}
and compute
\begin{equation}
 \mathrm{Im}\!\left[e^{-i\pi/4}\,f^{(\lambda)}(0)\right]
 = \frac{1}{\sqrt{2\pi\kappa}} \sum_{m=-\infty}^{+\infty} \mathrm{Im}\!\left[e^{-i\pi/4}\left(e^{2i\delta_{m}^{(\lambda)}} - 1\right)\right].
\end{equation}
Using the identity
\begin{equation}
 \mathrm{Im}\!\left[e^{-i\pi/4}\left(e^{2i\delta} - 1\right)\right]
 = \frac{1}{\sqrt{2}}\left[\sin 2\delta + 1 - \cos 2\delta\right]
 = \sqrt{2}\,\sin^{2}\delta,
\end{equation}
we find
\begin{equation}
 \mathrm{Im}\!\left[e^{-i\pi/4}\,f^{(\lambda)}(0)\right]
 = \frac{1}{\sqrt{\pi\kappa}} \sum_{m=-\infty}^{+\infty} \sin^{2}\delta_{m}^{(\lambda)}.
\end{equation}
Substituting into Eq.~\eqref{optical-theorem} yields
\begin{equation}
 \sigma_{\mathrm{tot}} = \sqrt{\frac{8\pi}{\kappa}} \cdot \frac{1}{\sqrt{\pi\kappa}} \sum_{m=-\infty}^{+\infty} \sin^{2}\delta_{m}^{(\lambda)}
 = \frac{4}{\kappa} \sum_{m=-\infty}^{+\infty} \sin^{2}\delta_{m}^{(\lambda)},
\end{equation}
which coincides with Eq.~\eqref{sigma-tot-sin2}, confirming the consistency of our results with the optical theorem.

\subsubsection*{Convergence and regularization}

For long-range potentials such as the Coulomb interaction, the partial-wave sum in Eq.~\eqref{sigma-tot-sin2} may diverge due to the slow decay of the phase shifts at large $|m|$. In the pure Coulomb problem, this divergence reflects the infinite range of the $1/r$ potential and leads to well-known subtleties in the definition of the total cross section.

In the helically twisted geometry, the effective Coulomb parameter is $\eta_{m} = -\omega k (m-\lambda)/\kappa$, which grows linearly with $|m|$. For large $|m|$, the geometric phase shift behaves as
\begin{equation}
 \delta_{m}^{(\mathrm{geom})} \sim \eta_{m}\left[\ln|\eta_{m}| - 1\right] + \mathcal{O}(1),
 \label{delta-large-m}
\end{equation}
which implies that $\sin^{2}\delta_{m}^{(\lambda)}$ does not vanish as $|m|\to\infty$. Consequently, the sum over partial waves in Eq.~\eqref{sigma-tot-sin2} is formally divergent.

This divergence is a standard feature of Coulomb-like scattering and can be handled by introducing a regularization procedure, such as screening the potential at large distances or restricting the sum to a finite number of partial waves $|m| \leq m_{\mathrm{max}}$. The regularized total cross section,
\begin{equation}
 \sigma_{\mathrm{tot}}^{(\mathrm{reg})} = \frac{4}{\kappa} \sum_{m=-m_{\mathrm{max}}}^{+m_{\mathrm{max}}} \sin^{2}\delta_{m}^{(\lambda)},
 \label{sigma-tot-reg}
\end{equation}
provides a finite and physically meaningful quantity that captures the scattering strength up to the cutoff angular momentum $m_{\mathrm{max}}$. The choice of $m_{\mathrm{max}}$ can be guided by experimental resolution or by the physical extent of the scattering region.

\begin{figure}[tbhp]
    \centering
    \includegraphics[width=0.48\textwidth]{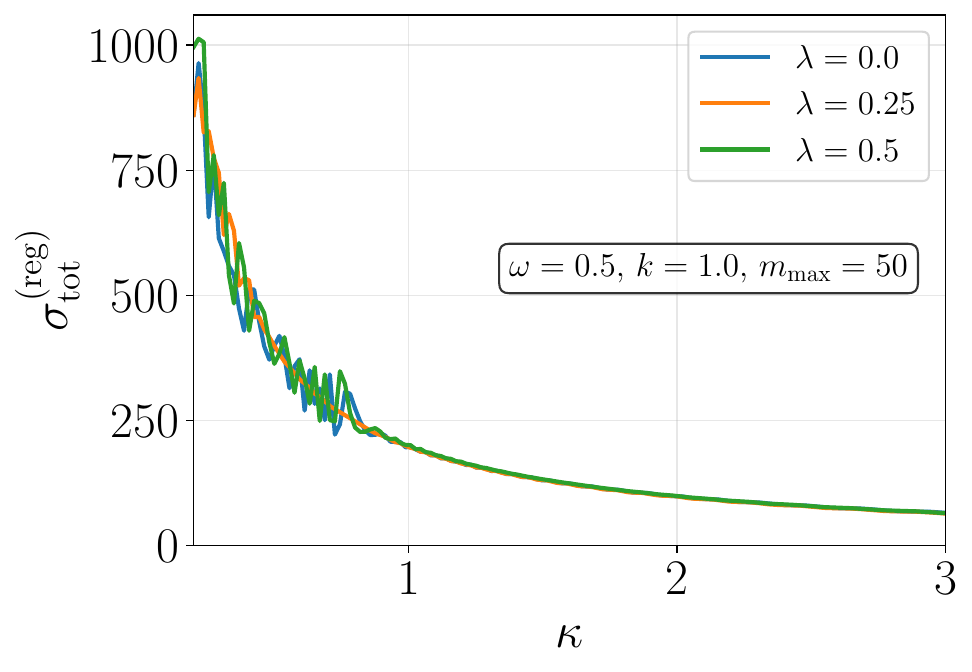}
    \caption{(Color online) Regularized total cross section $\sigma_{\mathrm{tot}}^{(\mathrm{reg})}$ as a function of the radial wave number $\kappa$ for the helically twisted Coulomb--Aharonov--Bohm problem. The curves are evaluated for $\omega = 0.5$, $k = 1$, and three values of the Aharonov--Bohm parameter, $\lambda = 0$ (blue), $\lambda = 0.25$ (orange), and $\lambda = 0.5$ (green), using a partial-wave cutoff $m_{\mathrm{max}} = 50$.}
    \label{fig:sigma_tot}
\end{figure}

Figure~\ref{fig:sigma_tot} displays the regularized total cross section $\sigma_{\mathrm{tot}}^{(\mathrm{reg})}$ as a function of the radial wave number $\kappa$ for fixed $\omega = 0.5$ and $k = 1$, with three values of the AB flux parameter. The total cross section exhibits the characteristic $1/\kappa$ scaling at high energies, which is a universal feature of two-dimensional scattering. At lower energies (small $\kappa$), the cross section increases and shows a stronger dependence on the AB flux, reflecting the enhanced role of the topological phase shifts in the threshold regime. The separation between curves at different $\lambda$ values demonstrates that the AB flux provides an effective control parameter for tuning the total scattering strength, complementing the geometric control provided by the torsion parameter $\omega$.

\subsection{Bound states from the poles of the partial-wave $S$ matrix}
\label{subsec:bound-from-poles}

Although our emphasis is on continuum scattering, the closed form of the partial-wave $S$ matrix provides a compact and fully equivalent route to the bound-state spectrum.
The key point is that the scattering solution in each angular channel is written in terms of Kummer's confluent hypergeometric function, and the corresponding $S$-matrix element can be expressed as a ratio of Gamma functions whose arguments are the same Kummer parameters.
As a consequence, bound states appear as poles of $S_{m}^{(\lambda)}$ after analytic continuation below the continuum threshold, and the pole condition reproduces the standard polynomial truncation condition of ${}_{1}F_{1}$.

Performing the analytic continuation $\kappa \to i\rho$ with $\rho>0$ as in Eq.~\eqref{kappa-to-rho}, which corresponds to energies below the continuum edge,
\begin{equation}
 E = E_{\infty}-\frac{\hbar^{2}\rho^{2}}{2\mu},
 \label{E-bound-param}
\end{equation}
where $E_{\infty}$ is given by Eq.~\eqref{E-infty}, the poles of $S_{m}^{(\lambda)}$ arise when the denominator Gamma function hits a non-positive integer. The pole condition, as derived in Eq.~\eqref{pole-condition-rho}, can be solved for the decay constant to yield
\begin{equation}
 \rho_{n,m}^{(\lambda)}
 =
 \frac{\omega k (m-\lambda)}
 {\,n+|m-\lambda|+\tfrac{1}{2}\,},
 \qquad n=0,1,2,\ldots,
 \label{rho-spectrum}
\end{equation}
and inserting this into Eq.~\eqref{E-bound-param} gives the bound-state energies in closed form:
\begin{equation}
 E_{n,m}^{(\lambda)}
 =
 -\frac{\hbar^{2}k^{2}}{2\mu}(1+\omega^{2})
 -\frac{\hbar^{2}}{2\mu}
 \left[
 \frac{\omega k (m-\lambda)}
 {\,n+|m-\lambda|+\tfrac{1}{2}\,}
 \right]^{2},
 \qquad n=0,1,2,\ldots
 \label{Enm-from-poles}
\end{equation}
The requirement $\rho_{n,m}^{(\lambda)}>0$ implies the existence condition $\omega k (m-\lambda)>0$, so that bound states occur only in the channels whose AB-shifted angular momentum has the same sign as $\omega k$.
Equation~\eqref{Enm-from-poles} is fully consistent with the bound-state quantization previously obtained in the helically twisted Coulomb-like problem without flux, and it reduces to that result upon setting $\lambda=0$.

To make explicit contact with the torsion-only bound spectrum reported in Ref.~\cite{AoP.2026.484.170295}, we now consider the vanishing Aharonov--Bohm flux limit of the bound-state energies obtained from the poles of the scattering matrix. Introducing the standard flux parameter $\phi \equiv e\Phi_B/(2\pi\hbar)$, the AB phase shifts the azimuthal index as $m\mapsto m_\phi \equiv m-\phi$ throughout the radial problem. Setting $\Phi_B=0$ (i.e., $\phi=0$ and $m_\phi\to m$), the pole condition in Eq.~\eqref{pole-condition-rho} reduces to
\begin{equation}
\frac{k\,\omega\, m\,\hbar}{\rho}=n+|m|+\frac{1}{2},
\label{eq:pole_quant_phi0}
\end{equation}
in exact correspondence with Eq.~(22) of Ref.~\cite{AoP.2026.484.170295}. Combining this with the relation \begin{equation}
\rho=\hbar^{-1}\sqrt{-2\mu E+\hbar^{2}k^{2}(1+\omega^{2})},
\label{eq:rho_relation}
\end{equation}
we obtain
\begin{equation}
E_{n,m}(\Phi_B=0)=
-\frac{\hbar^2 k^2}{2\mu}\,
\frac{4m^2+(1+2n)(1+\omega^2)(1+2n+4|m|)}{(1+2n+2|m|)^2},
\label{eq:Enm_phi0_eq23_form}
\end{equation}
which coincides exactly with Eq.~(23) of Ref.~\cite{AoP.2026.484.170295}.

Finally, since $\rho>0$, the quantization condition implies the usual existence condition for the attractive (binding) branch, namely $k\,\omega\,m>0$ (or, more generally, $k\,\omega\,m_\phi>0$ in the AB case), consistently with the torsion-induced Coulomb-like coupling in the effective potential.

\section{Conclusions \label{conclusion}}

We have presented an exact and unified treatment of bound--continuum physics for a charged quantum particle propagating in a helically twisted background, in the presence of an Aharonov--Bohm (AB) flux threading the twist axis. Starting from the stationary Schr\"odinger equation built with the Laplace--Beltrami operator of the helical metric, we showed that minimal coupling to the AB vector potential preserves separability and leads to a radial equation whose structure is identical to the canonical planar $2D$ Coulomb$+$AB problem. In this mapping, the topological flux enters exclusively through the gauge-invariant shift $m\to m-\lambda$, whereas the helical twist generates a Coulomb-like term proportional to $1/r$ with an effective strength controlled by the geometric parameter $\omega$ and the conserved longitudinal momentum $k$. In other words, the helical background realizes a \textit{geometry-programmable} Coulomb interaction whose sign and magnitude depend on the AB-shifted angular momentum channel.

In the scattering sector, the exact regular solutions were constructed in terms of Kummer confluent hypergeometric functions, enabling closed expressions for the partial-wave $S$ matrix and phase shifts. The resulting $S$ matrix factorizes into a purely AB contribution and a geometry-induced Coulomb factor expressed as a ratio of Gamma functions with argument $|m-\lambda|+\tfrac{1}{2}\pm i\eta_m$. This structure makes transparent the two distinct physical mechanisms at play: (i) a topological holonomy controlled by $\lambda$, and (ii) a long-range Coulomb phase controlled by the dimensionless parameter $\eta_m=\beta_{\mathrm{geom}}/\kappa=-\omega k (m-\lambda)/\kappa$, which governs the channel-dependent deformation of the phase shifts. Using the standard partial-wave reconstruction, we derived the scattering amplitude, the differential cross section, and the total cross section, demonstrating how the interplay between torsion-induced Coulomb phases and AB-induced angular-momentum shifts reshapes the angular distribution, producing flux-tunable interference patterns on top of the characteristic forward enhancement of long-range interactions.

The analysis of the phase shifts as a function of the torsion parameter $\omega$ revealed that the helical geometry provides a continuous control knob for the scattering observables. In particular, for the $s$-wave channel, the phase shift vanishes identically in the absence of AB flux but becomes activated when $\lambda \neq 0$, demonstrating a nontrivial interplay between topology and geometry. The total cross section, obtained both from direct partial-wave summation and verified via the optical theorem, exhibits the characteristic $1/\kappa$ scaling at high energies and shows a pronounced sensitivity to the AB flux in the threshold regime. The formal divergence of the total cross section, a hallmark of Coulomb-like scattering, was addressed through a regularization procedure that yields physically meaningful results.

Although our primary focus has been on the continuum regime, the analytic structure of the partial-wave $S$ matrix also provides a stringent consistency check and a compact bridge to the discrete spectrum. By analytic continuation $\kappa\to i\rho$, we showed that the pole condition of the Gamma functions reproduces the bound-state quantization rule previously obtained for the helically twisted Coulomb-like problem, now generalized by the replacement $m\to m-\lambda$. This confirms that the AB flux does not introduce additional bound levels by itself; rather, it reorganizes the existing geometry-induced spectrum through a topological relabeling of the angular channels and through the corresponding shift of the effective Coulomb parameter. In this way, the helically twisted Coulomb-AB system admits a fully consistent description in which bound states appear as $S$-matrix poles while scattering observables are encoded in the same closed-form phase shifts.

The present results highlight helically twisted backgrounds as analytically tractable platforms where geometry and topology can be combined to control quantum scattering without invoking \textit{ad hoc} model potentials. Several extensions are immediate. A natural next step is to incorporate spin degrees of freedom and possible spin--torsion or spin--orbit-like terms induced by the helical kinematics, which may generate additional channel splitting and polarization-dependent scattering. Relativistic generalizations of the Klein-Gordon or Dirac equations in the same geometry would clarify how the geometry-induced Coulomb coupling competes with relativistic thresholds and possible anomalous phases. Finally, it would be interesting to explore dynamical and transport implications, such as wave-packet scattering, time delays, and conductance-like quantities in helical waveguides, as well as synthetic implementations in photonic or cold-atom settings where both twist and AB-like phases can be engineered and tuned. These directions would further establish helical twisting as a versatile mechanism for geometry-controlled interference and long-range scattering in low-dimensional quantum systems.

\section*{Acknowledgement}

This work was partially supported by the Brazilian agencies CAPES, CNPq, and FAPEMA. E.O.S acknowledges the support from the grants CNPq/306308/2022-3, FAPEMA/UNIVERSAL-06395/22, CAPES/Finance Code 001.

\section*{\label{sec:datar}Data Availability Statement}

Data will be made available on reasonable request.

\end{document}